\documentclass[aps,prl,twocolumn,showpacs]{revtex4}
\usepackage{bm}
\usepackage{amsmath}
\usepackage{amssymb}
\usepackage{graphicx}

\newcommand{\beq}{\begin{equation}}
\newcommand{\eeq}{\end{equation}}
\newcommand{\beqar}{\begin{eqnarray*}}
\newcommand{\eeqar}{\end{eqnarray*}}

\newcommand{\Dh}{\hat{D}}
\newcommand{\Ch}{\hat{C}}
\newcommand{\Hh}{\hat{H}}
\newcommand{\Vh}{\hat{V}}

\newcommand{\Tc}{\mathcal{T}}

\newcommand{\Fc}{\mathcal{F}}
\newcommand{\Rc}{\mathcal{R}}
\newcommand{\Gc}{\mathcal{G}}
\newcommand{\Sc}{\mathcal{S}}
\newcommand{\Lc}{\mathcal{L}}

\newcommand{\pd}{\partial}
\newcommand{\rtarr}{\rightarrow}

\newcommand{\qb}{{\bf q}}

\newcommand{\Ab}{{\bf A}}
\newcommand{\rb}{{\bf r}}

\newcommand{\lan}{\langle}
\newcommand{\ran}{\rangle}
\newcommand{\om}{\omega}

\newcommand{\al}{\alpha}
\newcommand{\be}{\beta}
\newcommand{\ga}{\gamma}
\newcommand{\Ga}{\Gamma}
\newcommand{\de}{\delta}
\newcommand{\De}{\Delta}
\newcommand{\la}{\lambda}
\newcommand{\La}{\Lambda}
\newcommand{\sig}{\sigma}

\newcommand{\eps}{\varepsilon}

\newcommand{\lt}{\left}
\newcommand{\rt}{\right}
\newcommand{\RC}{\al_\ga}
\newcommand{\gaw}{\gamma_\text{w}}

\begin{document}

\title{Mesoscopic conductance fluctuations in graphene samples}
\author{Maxim Yu. Kharitonov$^{1}$ and Konstantin B. Efetov$^{1,2}$}
\affiliation{$^{1}$ Theoretische Physik III, Ruhr-Universit\"{a}t Bochum, Germany\\
$^{2}$L.D. Landau Institute for Theoretical Physics, Moscow, Russia}
\date{\today}

\begin{abstract}

Mesoscopic conductance fluctuations in graphene samples at
energies not very close to the Dirac point are studied
analytically. We demonstrate that the conductance variance $\langle
[\delta G]^2 \rangle$ is very sensitive to the elastic scattering
breaking the valley symmetry. In the absence of such scattering
(disorder potential smooth at atomic scales, trigonal warping
negligible), the variance $\langle [\delta G]^2 \rangle = 4 \langle [\delta G]^2
\rangle_\text{metal}$ is four times greater than  that in
conventional metals, which is due to the two-fold valley
degeneracy. In the absence of intervalley scattering, but for
strong intravalley scattering and/or strong warping $\langle [\delta
G]^2 \rangle  =2  \langle [\delta G]^2 \rangle_\text{metal}$. Only in the
limit of strong intervalley scattering $\langle [\delta G]^2 \rangle  =
\langle [\delta G]^2 \rangle_\text{metal}$. Our theory explains recent
numerical results and can be used for comparison with existing
experiments.

\end{abstract}
\maketitle

{\em Introduction.} Graphene (a monolayer of graphite) is a novel
material \cite{Novoselov,Zhang,Berger,Berger2} with the Dirac
electronic spectrum. A lot of progress in the theoretical
understanding of clean graphene has been made so far (see
e.g.~Ref. \cite{cleangraphene}). For many interesting effects,
however, disorder plays a significant role. A peculiar feature of
disordered graphene is that its physical properties are sensitive
to the scattering processes breaking the valley symmetry.

This sensitivity has been revealed in the behavior of the weak
localization (WL) correction to conductivity
\cite{Khvesh,WLFalko,Morpurgo,AlEf,WLFalkobilayer}. Another famous
phenomenon due to disorder that goes along with WL are the
mesoscopic conductance fluctuations (CF). CF with variance $\sim
e^2/\hbar$ are observed in graphene samples experimentally
\cite{Berger2,CFexp1,CFexp3,CFexp4}, although a detailed analysis
has not been reported.

Numerical investigation of CF in graphene has been undertaken
recently in Ref.~\cite{CFnum}. The authors have rather
unexpectedly found that CF in graphene were considerably stronger
than those in conventional metals~\cite{CFAlt,CFLeeStone} and the
variance did not seem to be universal. No clear explanation of
this effect was given in Ref.~\cite{CFnum}, although it was argued
that the unusual behavior might be due to percolation effects.

Here we develop an analytical theory of conductance fluctuations in
diffusive graphene samples at energies not very close to the Dirac
point. The results we obtain explain the findings of Ref.~\cite{CFnum} and
can be directly used for comparison with the experiments.

{\em Model.} We consider a general microscopic model of disorder
in graphene. The single-particle Hamiltonian of the system is
(we put $\hbar=1$ and recover it later on)
\beq
    \Hh_\rb = \Hh_0 + \Hh_\text{w} + \Vh(\rb), \mbox{ }
    \Hh_0=-i v ( 1^{KK'}\otimes \tau_\la^{AB}) \pd_\la. 
\label{eq:H} \eeq Here $\Hh_\text{w}=\mu_\text{w} \tau_z^{KK'}
\otimes [ \tau_x (\pd_x^2 -\pd_y^2) - 2\tau_y \pd_x \pd_y]^{AB} $
describes weak trigonal warping and $\la=x,y$. The Hamiltonian
$\Hh_\rb$ is a matrix in the tensor product $KK' \otimes AB$ of
the valley ($KK'$) and  sub-lattice ($AB$) spaces and $1$,
$\tau_{x,y,z}$ are the unity and Pauli matrices. The random
disorder potential $\Vh(\rb)$ is Gaussian with the correlation
function
\beq
    \lan \Vh(\rb) \otimes \Vh(\rb') \ran = 
        \{ \La_0 \openone \otimes \openone + \La^k_l \Tc_{kl} \otimes \Tc_{kl}\}
\delta(\rb-\rb'), \label{eq:VV} \eeq where $\openone = 1^{KK'}
\otimes 1^{AB}$, $\Tc_{kl}= \tau_k^{KK'}\otimes \tau_l^{AB}$,
$k,l=x,y,z$. For a given Fermi energy $\epsilon$ the scattering
rates (inverse scattering times) are defined as
\[
    (\ga_0,\ga_{zz}, \ga_{z\perp}, \ga_{\perp z}, \ga_{\perp \perp})\equiv \pi \nu
    (\La_0,\La^z_z, \La^z_{x,y}, \La^{x,y}_z, \La^{x,y}_{x,y}),
\]
where $\nu=\epsilon/(2 \pi v^2)$ is the density of states per one
valley and one spin.

In Eq.~(\ref{eq:VV}), the term $\propto \ga_0$ arises from remote charge impurities
in the substrate,
the field of which
varies smoothly  at atomic scales,
while the rest of the terms
describe various atomically-sharp defects that break the valley symmetry.
Being diagonal in $KK'$-space ($\propto~\tau_z^{KK'}$), the terms
$\propto \ga_{zz}$ and $\propto\ga_{z\perp}$ do not involve the
intervalley scattering, but do lift the valley degeneracy by
acting differently on the valleys. Such terms describe the
intravalley scattering, whereas the terms $\propto \ga_{\perp z}$
and $\propto \ga_{\perp\perp}$ are due to

\begin{figure}
\includegraphics{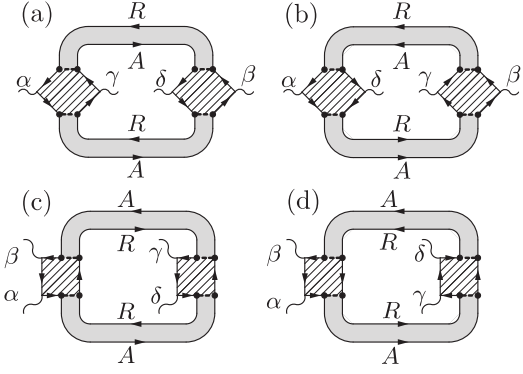}
\caption{Diagrams for the conductivity correlation function
$F_{\al\be,\ga\de}(\De \epsilon, H, \De H)$
[Eq.~(\ref{eq:F})]. Gray stripes denote diffusons and Cooperons, rendered with lines blocks are Hikami boxes, see Fig.~\ref{fig:blocks}.
The diagrams (c),(d)
with the substitution $\al \leftrightarrow \be$, $\ga \leftrightarrow \de$,
$R \leftrightarrow A$ must also be considered.}
\label{fig:CFs}
\end{figure}

{\em Calculations.}
We calculate the correlation function
\beq
    F_{\al\be,\ga\de}(\De \epsilon, H,\De H) =
        \lan \de \sig_{\al\be}(\epsilon+\De\epsilon, H+\De H)
           \de \sig_{\ga\de}(\epsilon, H) \ran
\label{eq:Fdef}
\eeq
of the conductivities $\sig_{\al\be}(\epsilon+\De\epsilon, H+\De H)$
and $\sig_{\ga\de}(\epsilon, H)$ taken at  the Fermi energies
$\epsilon+\De\epsilon$, $\epsilon$ \cite{mesoscale} and
magnetic fields  $H+\De H$, $H$
($\al,\be,\ga,\de=x,y$ and $\de \sig = \sig - \lan \sig \ran$).

We use the averaging technique developed for conventional
disordered metals \cite{AGD}. We assume (i) weak disorder,
$\La_0/v^2 \ll 1$, and (ii) diffusive regime, i.e., that the mean
free path $l=v/\ga_0$ is much smaller than the size of the sample
and the valley-symmetric rate $\ga_0$ is dominant, $\ga_0 \gg
\ga_{\{z,\perp\},\{z,\perp\}}$. As it was shown in
Ref.~\cite{AlEf} one should first renormalize the velocity $v$ and
constants $\La $ [Eq.~(\ref{eq:VV})] solving renormalization group
equations and then use them for calculating the localization
corrections. This procedure is valid so long as $\epsilon \gtrsim
\epsilon_0\exp(-\pi v^2/\La_0) $, where $\epsilon_0$ is an
atomic-scale energy. The same can be done for the correlation
function (\ref{eq:Fdef}) and we further assume that $v$ and
$\La$'s have been renormalized. Under these assumptions, the
calculations for graphene generalize those for ordinary
metals~\cite{CFAlt,CFLeeStone} and $F_{\alpha \beta ,\gamma \delta
}(\Delta \epsilon,H,\Delta H)$ is given by the diagrams in
Fig.~\ref{fig:CFs}.

\begin{figure}
\includegraphics{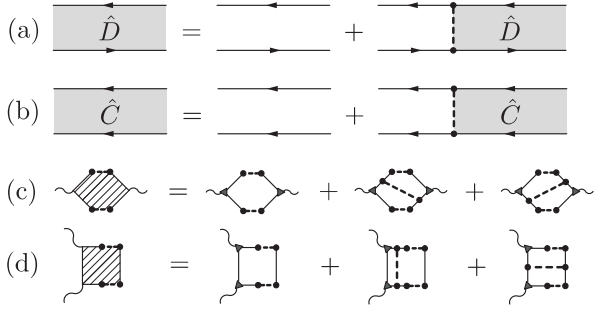}
\caption{ (a),(b) Diagrammatic representation of the integral equations for the diffuson and Cooperon. (c),(d) Hikami boxes. The current vertex
renormalized by disorder (dark triangle) equals $\tilde{j}_\al=2e v (1^{KK'} \otimes \tau_\al^{AB})$.}
\label{fig:blocks}
\end{figure}

The arising disorder-averaged products of the exact retarded ($R$)
and advanced ($A$) Green's functions $\hat{\Gc}^{R,A}$ yield the
diffusons and Cooperons defined as:
\begin{eqnarray}
    \Dh_\om(\rb,\rb') \equiv \lan \hat{\Gc}^R (\eps+\om,\rb,\rb')
    \otimes \hat{\Gc}^A (\eps,\rb',\rb) \ran, \label{eq:Ddef}
\\
    \Ch_\om(\rb,\rb') \equiv \lan \hat{\Gc}^R (\eps+\om,\rb,\rb')
    \otimes \hat{\Gc}^A (\eps,\rb,\rb') \ran. \label{eq:Cdef}
\end{eqnarray}
They satisfy the integral equations represented diagrammatically
in Fig.~\ref{fig:blocks}(a),(b). These equations possess a
nontrivial matrix structure acquired from the correlator
(\ref{eq:VV}) (dashed lines) and disorder-averaged $\lan
\hat{\Gc}^{R,A} \ran$ (fermionic lines). Consequently, there exist
``high-energy'' modes of $\Dh$ and $\Ch$ with gaps $\sim \ga_0$ as
well as ``low-energy'' modes, which gaps do not contain $\ga_0$
\cite{WLFalko,AlEf}. Giving much greater contribution in the
diffusive regime, only the latter low-energy modes are of interest
here.

The effect of trigonal warping can be taken into account in $\hat{\Gc}^{R,A}$
up to  the second order in $\Hh_\text{w}$.
As a result, the self-energies of the  Green's functions $\lan\hat{\Gc}^{R,A}\ran$
acquire the warping rate $\ga_\text{w} = ( \mu_\text{w} \epsilon/v^2)^2/\ga_0$
and the dashed line in Fig.~\ref{fig:blocks}(a),(b),
in addition to the correlator (\ref{eq:VV}),
also represents the ``warping term''
\[ \hat{\ga}_\text{w}^{D,C}=\pm \gaw (\tau_z^{KK'} \otimes 1^{AB})_R
    \otimes (\tau_z^{KK'} \otimes 1^{AB})_A
\]
with $+$ and $-$ for the diffuson [$D$,(a)] and Cooperon
[$C$,(b)], respectively. It appears that, in the {\em low-energy}
diffuson/Cooperon subspaces, the  matrix structure of the warping
term $\hat{\ga}_\text{w}^{D,C}$  is {\em identical} to that of the
intravalley scattering of type $\hat{\ga}_{zz} = \ga_{zz} \Tc_{zz}
\otimes \Tc_{zz}$ in Eq.~(\ref{eq:VV}). Therefore, the effect of
warping on the low-energy diffusion modes of $\Dh$ and $\Ch$ is
not any different from that of the intravalley scattering $\propto
\ga_{zz}$ and the effect of warping could be taken into account by
the substitution $\ga_{zz} \rtarr \ga_{zz} +\gaw$.

Resolving the matrix-structure of the equations in
Fig.~\ref{fig:blocks}(a),(b), we obtain:
\begin{widetext}
\begin{eqnarray}
    \Dh_\om(\rb,\rb') = \frac{\pi \nu}{16} \lt\{\lt[D^0_\om+D^1_\om+2 D^{2,3}_\om\rt]
                            1\otimes 1+
    \lt[D^0_\om+D^1_\om-2 D^{2,3}_\om\rt] \tau_z\otimes \tau_z +
    [D^0_\om-D^1_\om] \tau_\la\otimes \tau_\la \rt\}\otimes
       (1 \otimes 1+\tau_k \otimes \tau_k)
\label{eq:D}
\\
    \Ch_\om(\rb,\rb') = \frac{\pi \nu}{16}
\lt\{\lt[ 2 C^{2,3}_\om+ C^1_\om+C^0_\om \rt]
                            1\otimes 1+
    \lt[ 2C^{2,3}_\om- C^1_\om-C^0_\om\rt] \tau_z\otimes \tau_z +
    [C^1_\om-C^0_\om]  \tau_\la\otimes \tau_\la \rt\} \otimes
        (1 \otimes 1-\tau_k \otimes \tau_k)
\label{eq:C}
\end{eqnarray}
\end{widetext}
In Eqs.~(\ref{eq:D}) and (\ref{eq:C}), the
tensor products are ordered 
as $(R\otimes A)_{KK'}\otimes(R\otimes A)_{AB}$, $\la=x,y$, and
$k=x,y,z$. The diffuson/Cooperon components $D^i_\om =D^i_\om
(\rb,\rb')$ and $C^i_\om =C^i_\om (\rb,\rb')$, $i=0,1,2,3$,
satisfy the equations
\beq
    \{-i\om -D\nabla_{D,C}^2+\Ga_i +
    \ga_\text{inel}\} (D,C)^i_\om(\rb,\rb') = \de(\rb-\rb'),
\label{eq:DCdiffeq} \eeq where $ \nabla_D=\nabla-i(e/c)\De
\Ab(\rb)$, $\nabla_C= \nabla-i(e/c)[2 \Ab(\rb)+\De \Ab(\rb)]$, the
vector potentials $ \Ab(\rb)$, $\De \Ab(\rb)$ correspond to $H$
and $\De H$, respectively, $D=v^2 /\ga_0$ is the diffusion
coefficient, and $\ga_\text{inel}$ is the inelastic scattering
rate due to, e.g., electron-electron or electron-phonon
interactions. The elastic scattering rates $\Ga_i$ due to disorder
equal
\beq
    \Ga_0 = 0, \mbox{ } \Ga_1 = 4 \ga_\perp, \mbox{ }
    \Ga_2=\Ga_3= 2 \ga_\perp + 2 \ga_z + 2 \gaw,
\label{eq:Ga} \eeq where the total intervalley $ \ga_\perp=
\ga_{\perp z} + 2\ga_{\perp\perp}$ and intravalley $\ga_z=
\ga_{zz} + 2 \ga_{z\perp}$ scattering rates were introduced.

The Cooperon in the form of Eq.~(\ref{eq:C}) has been obtained
earlier~\cite{WLFalko,AlEf}, whereas the form (\ref{eq:D}) of the
diffuson is obtained here for the first time. Note that for a
given $i$ the rates $\Ga_i$ [Eq.~(\ref{eq:Ga})] entering the
corresponding
diffuson $D_\om^i$ and Cooperon $C_\om^i$ modes 
are identical. The fact that  the insensitive to various
phase-breaking phenomena diffuson $\Dh$ does contain the rates
$\ga_z$, $\ga_\perp$,  and $\gaw$~\cite{nn} means that the effects
of the intravalley and intervalley scattering and of the trigonal
warping should never be understood as a suppression of electron
interference alone.

{\em Results.} Calculating the diagrams in Fig.~\ref{fig:CFs}, for
the correlation function of {\em
conductivities}~[Eq.~(\ref{eq:Fdef})] we obtain
\begin{widetext}
\begin{eqnarray}
    F_{\al\be,\ga\de}(\De \epsilon, H,\De H) = \lan \de \sig_{\al\be}(\epsilon+\De\epsilon, H+\De H)
    \de \sig_{\ga\de}(\epsilon, H)
    \ran=
    ( 2_s \, 2_v e^2 D )^2 \int \frac{d\eps d\eps'}{(2\pi)^2}
    \frac{d n(\eps)}{d\eps} \frac{d n(\eps')}{d\eps'}
      \frac{1}{\Sc^2} \nonumber\\
\times \int d\rb\, d\rb'\, \sum_{i=0}^{3}\lt\{
        \de_{\al\ga} \de_{\be\de} |D^{i}_\om(\rb,\rb')|^2+
        \de_{\al\de} \de_{\be\ga} |C^i_\om(\rb,\rb')|^2
        +\frac{1}{2} \de_{\al\be} \de_{\ga\de}  \text{Re} [ D^i_\om(\rb,\rb')D^i_\om(\rb',\rb)
            + C^i_\om(\rb,\rb') C^i_\om(\rb',\rb)]
    \rt\},
\label{eq:F}
\end{eqnarray}
\end{widetext}
where $\om=\eps-\eps'+\De\epsilon$, $\Sc $ is the sample area,
$n(\eps)=1/[\exp(\eps/T)+1]$ is the Fermi distribution function,
and the factors $2_s$ and $2_v$ originate from the dimensionality
of the spin and valley spaces
(the indices $s$ and $v$ emphasize their origin).

Equation~(\ref{eq:F}),
together with Eqs.~(\ref{eq:D})-(\ref{eq:Ga}),
constitutes the main result of our work.
The key feature 
characterizing graphene is that different diffuson and Cooperon
modes $i$ enter Eq.~(\ref{eq:F}). The magnitude of mesoscopic
fluctuations is thus determined by the strength of elastic
scattering processes breaking the valley symmetry. When all such
effects are negligible, the result~(\ref{eq:F}) for graphene is
$(2_v)^2=4$ times greater than that for conventional
metals~\cite{CFAlt,CFLeeStone} due to the two-fold valley
degeneracy.
Note that at $H=\De H=0$ one has $D_\om^i=C_\om^i$ and for a given $i$
the diffuson
and Cooperon
contribute equally.

To be specific, below we consider the case of a rectangular sample
with length $\Lc_x$ and width $\Lc_y$, occupying the area
$0<x<\Lc_x$, $0<y<\Lc_y$, and attached to ideal leads at $x=0$ and
$x=\Lc_x$. The conductance $G=G_{xx}$ in the $x$ direction is
related to the conductivity $\sig_{xx}$ as $G=\sig_{xx}
\Lc_y/\Lc_x$. From Eq.~(\ref{eq:F}), at $T, \ga_\text{inel} \ll
\epsilon^*_x$, where $\epsilon_x^*=\pi^2 D /\Lc_x^2$ is the
Thouless energy for the $x$ dimension, for the {\em conductance}
correlation function $
    \Fc (\De \epsilon )=\lan \de G(\epsilon+\De \epsilon)
\de G(\epsilon) \ran = (\Lc_y/\Lc_x)^2 F_{xx,xx}(\De  \epsilon,H,0)
$ we obtain
\begin{eqnarray}
    \Fc(\De \epsilon) &=& \al_H\lt[\frac{ 2_s \, 2_v e^2 D}{2 \pi}\rt]^2
    \frac{1}{\Lc_x^4} \nonumber
    \\ &  \times &
    \sum_{i=0}^{3} \sum_\qb \lt\{ 2 |D^i_{\De \epsilon}(\qb)|^2
         +\text{Re} [D^i_{\De \epsilon}(\qb)]^2 \rt\},
\label{eq:Fceps}
\end{eqnarray}
where $D_{\De \epsilon}^i(\qb) = 1/(-i\De \epsilon  +D\qb^2
+\Ga_i)$ are the spatial eigenmodes of the diffuson,
$\qb^2=q_x^2+q_y^2$, $q_x= \pi n_x/\Lc_x$, $n_x=1,2,\ldots,$ and
$q_y= \pi n_y/\Lc_y$, $n_y=0,1, \ldots.$ In Eq.~(\ref{eq:Fceps}),
the factor $\al_H$ accounts for the sensitivity of the Cooperons
to the magnetic field in the two limiting cases: $\al_H=1$ for $H
\ll H^*$ and $\al_H=1/2$ for $H \gg H^*$, where $H^*=(c/e)/
\Lc_x^2$. The conductance variance $\Fc(\De \epsilon=0)=\lan [\de
G]^2 \ran$ following from Eq.~(\ref{eq:Fceps}) equals
\begin{eqnarray}
    \lan [\de G ]^2 \ran = 3 \al_H \lt[\frac{ 2_s \, 2_v e^2 }{2 \pi \hbar}\rt]^2
     \sum_{i=0}^{3}  \Rc(\Lc_i,\Lc_x,\Lc_y),
\label{eq:dG2}
\\
    \Rc(\Lc_i,\Lc_x,\Lc_y)
=\frac{1}{\pi^4 \Lc_x^4}
          \sum_{n_x=1}^{\infty} \sum_{n_y=0}^{\infty}
    \lt[\frac{1}{\Lc_i^2} +\frac{n_x^2}{\Lc_x^2} +\frac{n_y^2}{\Lc_y^2}\rt]^{-2},
\label{eq:R}
\end{eqnarray}
where $\Lc_i^2 = \pi^2 D /\Ga_i$. For both narrow ($\Lc_y \ll
\Lc_x$) and wide ($\Lc_y \gg \Lc_x$) samples, the contribution of
a given mode $i$ to Eq.~(\ref{eq:R}) is unsuppressed if $\Lc_i \gg
\Lc_x$ and equals:
\[
    \Rc_0=\Rc(\infty,\Lc_x,\Lc_y)=
    \lt\{ \begin{array}{ll} 1/90,  & \Lc_y \ll \Lc_x , \\
        \zeta(3) \Lc_y /(4\pi^3 \Lc_x),& \Lc_x\ll\Lc_y.
        \end{array} \rt.
\]
Wide samples are thus more attractive for the observation of
unsuppressed CF. In this case the length $\Lc_i$ has to be greater
than only the shorter dimension $\Lc_x$ (equivalently, $\Ga_i \ll
\epsilon^*_x$), but can be arbitrary compared to $\Lc_y$.

The limiting cases of Eq.~(\ref{eq:dG2}) for different strengths
of the scattering processes can be summarized as follows:
\beq
    \lan [\de G ]^2 \ran  = \frac{2_v^2}{4} \RC  \lan [\de G ]^2 \ran_\text{m}, \mbox{ }
    \lan [\de G ]^2 \ran_\text{m}=12\al_H \lt[\frac{ 2_s e^2 }{2 \pi \hbar}\rt]^2  \Rc_0,
\label{eq:dG2lim} \eeq where the conductance variance $\lan [\de
G]^2 \ran_\text{m}$ for a conventional metal
\cite{CFAlt,CFLeeStone} was introduced. The coefficient $\RC$
gives the number of diffusion modes $i$ that contribute (i.e., for
which $\Ga_i \ll \epsilon_x^*$) to CF, see Table~\ref{tab:RC}. As
follows from Eq.~(\ref{eq:Ga}), the mode $i=0$ (``pseudo-spin
singlet'') is unaffected by any of the scattering mechanisms, the
mode $i=1$ (``triplet, 0'') can be suppressed by the intervalley
scattering only, and the modes $i=2,3$ (``triplet, $\pm 1$'') can
be suppressed by both intervalley and intravalley scattering and
by trigonal warping. Note that trigonal warping  does affect CF,
in the same way as intravalley scattering does.

(i)~When all the effects are negligible, $\ga_\perp, \ga_z,
\gaw\ll \epsilon^*_x$, all four modes contribute equally, $\RC =4 $,
and $\lan [\de G ]^2 \ran =4\lan [\de G ]^2 \ran_\text{m}$
is four times greater than that for a
conventional metal. This is explained by an additional
two-fold valley degeneracy described by the factor $2_v$ in
Eq.~(\ref{eq:dG2lim}).

(ii)~If the intervalley scattering is weak, $ \ga_\perp \ll \epsilon^*_x$,  but
either the intravalley scattering or the trigonal warping are sufficiently strong,
$\ga_z \gg \epsilon^*_x$ or $\gaw \gg \epsilon^*_x$, then
the two modes $i=0,1$ contribute, while the modes $i=2,3$ are suppressed.
In this case $\RC=2$ and $\lan [\de G ]^2 \ran =2\lan [\de G ]^2 \ran_\text{m}$
is two times greater than that for a metal.

(iii)~Finally, if the intervalley scattering is strong, $ \ga_\perp \ll \epsilon^*_x$,
and the intravalley scattering $\ga_z$ and trigonal warping $\gaw$  rates
are arbitrary compared to $\epsilon^*_x$,
then all triplet modes $i=1,2,3$ are suppressed, and only the gapless mode $i=0$ contributes.
In this case $\RC=1$ and $\lan [\de G ]^2 \ran =\lan [\de G ]^2 \ran_\text{m}$
coincides with that
for a metal.

\begin{table}
\begin{tabular}{|c|c|c|}
    \hline
       & $\ga_z \ll \epsilon^*_x$ and $\ga_\text{w} \ll \epsilon^*_x$ &  $\ga_z \gg \epsilon^*_x$ or $\ga_\text{w} \gg \epsilon^*_x$\\
    \hline
    $\ga_\perp \ll \epsilon^*_x$ & 4 & 2  \\
    $\ga_\perp \gg \epsilon^*_x$ & 1 & 1   \\
    \hline
\end{tabular}
\caption{The number $\RC$ of the diffusion modes contributing
to the conductance variance in graphene
for different intervalley $\ga_\perp$,
intravalley $\ga_z$ and trigonal warping $\ga_\text{w}$ scattering rates.
The value of $\RC$ also gives the ratio of the conductance variance in graphene to that in
conventional metal, $\lan [\de G]^2 \ran_\text{graphene}
        =\RC \lan [\de G]^2 \ran_\text{metal}$, see Eq.~(\ref{eq:dG2lim}).}
\label{tab:RC}
\end{table}

The conductances fluctuations in graphene were studied numerically
in Ref.~\cite{CFnum} (see Fig.~3 therein). For atomically-sharp
disorder, a plateau $ \lan [\de G ]^2 \ran_\text{plateau} \approx
\lan [\de G]^2 \ran _\text{m} $ in the dependence of $\lan [\de G
]^2 \ran $ on the disorder strength $\La_0 /v^2$ was obtained,
which clearly corresponds to the case (iii). For atomically-smooth
disorder, a wide peak in the dependence of $\lan [\de G ]^2 \ran$
on 
$\La_0 /v^2$
with maximum
$\lan [\de G ]^2 \ran_\text{max} \approx (4.5 - 5)\lan [\de G ]^2 \ran _\text{m} $
at $\La_0 /v^2 \sim 1$ was obtained.
We believe this situation corresponds to the case (i),
the maximum value being close to our prediction.
We emphasize that our theory,
just like that of Refs.~\cite{CFAlt,CFLeeStone},
requires {\em both}  weak disorder ($\La_0 /v^2 \ll 1$)
and  the diffusive regime [$l=v/\ga_0 \ll \Lc=\min(\Lc_x,\Lc_y)$].
The reason for having a peak, rather than a plateau, for smooth disorder
in Ref.~\cite{CFnum} is that the range of $\La_0/v^2$, where both these conditions
are met, is quite narrow.
The diffusive regime 
is not reached until disorder becomes strong ($\La_0/v^2 \gtrsim
1$), while for smaller values of $\La_0/v^2\ll 1$ the system is
simply in the ballistic regime $l \gtrsim \Lc$. This is supported
by direct check of parameters ($l/\Lc \sim 0.1$ for $\La_0/v^2
\sim 1$ and thus $l/\Lc \sim 1$ for $\La_0/v^2 \sim 0.1$ ) and by
an improving tendency (earlier upsurge of $\lan [\de G]^2\ran$
with increasing $\La_0/v^2 \ll 1$) for larger  samples (filled vs.
open symbols).

Our theory thus helps understand the findings of Ref.~\cite{CFnum}
without assuming the existence of percolation paths and
nonergodicity as was done  by the authors. The ergodicity implies
equivalence of averaging over disorder and  the Fermi energy (or
magnetic field), $
    \lan f[G(\epsilon)] \ran \doteq \lim_{\De \epsilon \rtarr \infty}
    \frac{1}{\De\epsilon}
        \int_{\epsilon-\De \epsilon/2}^{\epsilon+\De \epsilon/2} d\epsilon' f[G(\epsilon')].
$ For $f[G(\epsilon)]=G(\epsilon)$, this is clearly true, if
$\Fc(\De\epsilon) \rtarr 0 $ as $\De\epsilon \rtarr \infty$, see
Eq.~(\ref{eq:Fceps}). This asymptotic of $\Fc(\De\epsilon) $ is
determined by the behavior of the diffusion modes $D^i_{\De
\epsilon}(\qb)$ at large energies $\De \epsilon$ and, in this
respect, graphene is not any different from an ordinary metal
\cite{CFAKL}. One can estimate $
    \Fc(\De\epsilon) \propto
\int^{+\infty}_{\sqrt{\De\epsilon/D}} q\, dq \, \frac{1}{q^4}
\propto \frac{D}{\De\epsilon}
$ for $\De \epsilon \gg \epsilon_x^*, \epsilon_y^*$. The proof for
$f[G(\epsilon)]=[G(\epsilon)]^m$, $m>1$, is analogous. Thus, the
ergodic hypothesis for graphene holds. The violation of ergodicity
in Ref.~\cite{CFnum} occured near the Anderson metal-insulator
transition, and might be due to the fact  that averaging over
energy was mixing extended and localized states.

{\em Conclusion.} We have developed a theory of conductance
fluctuations in monolayer graphene samples. We expect our findings
presented in the {\em Results} section to be also completely
applicable to bilayer graphene samples.

We thank SFB Transregio 12 for financial support.


\begin{thebibliography}{99}
\bibitem{nn}
However, the density-density correlation function
$
    P_\om(\rb,\rb')=2_s \, 2_v \nu [\de(\rb-\rb')+i\om D_\om^0(\rb,\rb')]
$, obtained from Eq.~(\ref{eq:D}),
is expressed solely through the gapless mode $D_\om^0$, as one would expect from the
particle conservation law.
\end{thebibliography}
\end{document}